\documentclass[a4paper,12pt,english,twoside]{article}
\usepackage{graphicx,geometry,babel,amssymb,amstext,amsmath}
\usepackage[T1]{fontenc}
\usepackage[latin9]{inputenc}

\title{Quantum Discord in the Ground and Thermal States of Spin Clusters}
\author{Amit Kumar Pal and Indrani Bose\footnote{indrani@bosemain.boseinst.ac.in}\\
       Department of Physics\\
       Bose Institute\\
       93/1 A. P. C. Road, Kolkata-700009, India}
\date{}

\begin{document}

\maketitle

\begin{abstract}
Quantum discord is a general measure of bipartite quantum correlations with a potential role in quantum 
information processing tasks. Spin clusters serve as ideal candidates for the implementation of some of the 
associated protocols. In this paper, we consider a symmetric spin trimer and a tetramer which describe a 
number of known molecular magnets and compute the quantum discord in the ground and thermal states of the 
clusters. The variations of the quantum discord as a function of an anisotropy parameter, magnetic field and 
temperature are investigated. We obtain a number of interesting results such as a finite value of the quantum 
discord in the trimer ground state for which the pairwise entanglement is known to be zero, differences in the 
nature of some of the variations in the ferromagnetic and antiferromagnetic cases and discontinuous jumps in 
the magnitude of the quantum discord at first order quantum phase transition points. A remarkable feature that 
is observed is that the quantum discord completely vanishes only in the asymptotic limit of temperature 
$T\rightarrow\infty$.
We further study the dynamics of the quantum discord and the pairwise entanglement at $T=0$ under the effect 
of a dephasing channel describing the interaction of the reduced spin cluster state with independent local 
environments. The QD is found to vanish asymptotically as $t\rightarrow\infty$. 
In the case of the spin trimer, the pairwise entanglement has a zero value at all times and reaches 
a zero value in a finite time in the case of the tetramer.      
\end{abstract}


\bigskip\bigskip

\section{Introduction}

An interacting quantum system is characterized by the presence of
correlations amongst its different constituents. The correlations
have, in general, both classical and quantum components. The most
prominent example of quantum correlations is that of entanglement
which serves as the fundamental resource in several quantum information
processing tasks such as quantum computation, teleportation and dense
coding \cite{nielsen}. Entanglement can be of various types, e.g., bipartite,
multipartite, zero temperature, finite temperature etc. for which a number
of quantitative measures exist. The entanglement content of quantum states 
and its variation as a function of changing parameters have been extensively 
investigated  in recent times \cite{vedralRMP,senAP,latorre}.
A different measure of pairwise quantum correlations, namely, the quantum
discord (QD) has further been proposed based on the information-theoretic
concept of mutual information \cite{olivier,zurekRMP,henderson,sarandy,luo}. 
The basic difference of the QD from entanglement
is evident from the fact that the QD is non-zero in certain separable
states which are, by definition, unentangled. The QD is defined to be the difference
between two quantum extensions of the classical mutual information.
In the classical domain, the two representations are exactly equivalent. 

In classical information theory, the correlation between two random
variables $A$ and $B$ is measured by their mutual information \cite{nielsen}
\begin{equation}
I(A,B)=H(A)+H(B)-H(A,B)
\label{mutual1}
\end{equation}
The random variable $A$ takes on the values \textquoteleft$a$\textquoteright \,  
with probabilities given by the set $\{p_{a}\}$. $H(A)=-\sum_{a}p_{a}\log_{2}p_{a}$
is the Shannon entropy. $H(A,B)$ corresponds to the joint Shannon entropy
defined as $H(A,B)=-\sum_{a,b}p_{a,b}\log_{2}p_{a,b}$. As alternative
representation of the mutual information is given by \cite{sarandy,luo}
\begin{equation}
J(A,B)=H(A)-H(A|B)
\label{mutual2}
\end{equation}
where $H(A|B)$ is the conditional entropy and quantifies the lack
of knowledge of the value of $A$ when that of $B$ is known. The
exact equivalence of the expressions in equations (\ref{mutual1}) and (\ref{mutual2}) can be demonstrated
using the Bayes' rule \cite{nielsen}. 

The generalisation of the expressions to the quantum case is achieved via
the replacement of the classical probability distribution and the
Shannon entropy by the density matrix $\rho$ and the Von Neumann entropy,
$S(\rho)=-tr\left(\rho\log_{2}\rho\right)$, respectively. Thus, the
quantum versions of equations (\ref{mutual1}) and (\ref{mutual2}) can be written as 
\begin{equation}
I\left(\rho_{AB}\right)=S\left(\rho_{A}\right)+S\left(\rho_{B}\right)-S\left(\rho_{AB}\right)
\label{mutual3}
\end{equation}
\begin{equation}
J\left(\rho_{AB}\right)=S\left(\rho_{A}\right)-S\left(\rho_{A}\left|\rho_{B}\right.\right)
\label{mutual4}
\end{equation}
where $S\left(\rho_{AB}\right)$ is the quantum joint entropy and
$S\left(\rho_{A}\left|\rho_{B}\right.\right)$ the quantum conditional
entropy. The latter quantity is, however, not properly defined by
a simple replacement of the Shannon entropy by the Von Neumann entropy.
The magnitude of the quantum conditional entropy, by the very nature
of its definition (ignorance of $A$ once $B$ is known), depends
on the type of measurement. Since different measurement choices
yield different results, equations (\ref{mutual3}) and (\ref{mutual4}) are no longer identical.
We consider Von Neumann-type measurements on $B$ defined in terms
of a complete set of orthogonal projectors $\left\{ \Pi_{i}\right\} $,
corresponding to the set of possible outcomes $i$. The state of the system, once the measurement
is made, is given by 
\begin{equation}
\rho_{i}=\left(I\otimes\Pi_{i}^{B}\right)\rho_{AB}\left(I\otimes\Pi_{i}^{B}\right)/p_{i}
\end{equation}
with 
\begin{equation}
p_{i}=tr \left(\left(I\otimes\Pi_{i}^{B}\right)\rho_{AB}\left(I\otimes\Pi_{i}^{B}\right)\right)
\end{equation}
$I$ denotes the identity operator for the subsystem $A$ and $p_{i}$
gives the probability of obtaining the outcome $i$. From equation \ref{mutual4},
an alternative expression of quantum mutual information is given by \cite{sarandy,luo}
\begin{equation}
J\left(\rho_{AB},\left\{ \Pi_{i}^{B}\right\} \right)
=S\left(\rho_{A}\right)-S\left(\rho_{AB}\left|\left\{ \Pi_{i}^{B}\right\} \right.\right)
\end{equation}
The quantum analog of the conditional entropy is 
\begin{equation}
S\left(\rho_{AB}\left|\left\{ \Pi_{i}^{B}\right\} \right.\right)
=\sum_{i}p_{i}S\left(\rho_{i}\right)
\end{equation}
Henderson and Vedral \cite{henderson} have shown that the maximum of 
$J\left(\rho_{AB},\left\{ \Pi_{i}^{B}\right\} \right)$
w.r.t. $\left\{ \Pi_{i}\right\} $ provides a measure of the classical
correlations, $C\left(\rho_{AB}\right)$, i.e., 
\begin{equation}
C\left(\rho_{AB}\right)
=\underset{\left\{ \Pi_{i}^{B}\right\} }{\max}
\left(J\left(\rho_{AB},\left\{ \Pi_{i}^{B}\right\} \right)\right)
\label{class}
\end{equation}
The difference between the total correlations $I\left(\rho_{AB}\right)$
(equation \ref{mutual3}) and the classical correlations $C\left(\rho_{AB}\right)$
defines the QD, $Q\left(\rho_{AB}\right)$, 
\begin{equation}
Q\left(\rho_{AB}\right)=I\left(\rho_{AB}\right)-C\left(\rho_{AB}\right)
\end{equation}
In the case of pure states, one can show that the QD reduces to  the entropy
of entanglement \cite{luo} so that entanglement provides the sole contribution
to quantum correlations. In the case of mixed states, however, the QD
and entanglement provide different measures of quantum correlations.
QD has been quantified in a number of two qubit states
\cite{sarandy,luo,ali,maziero1,werlang1}. For spin Hamiltonians with certain symmetries, the two-spin
reduced density matrix $\rho_{ij}$ (the two spins are located at
the sites $i$ and $j$) can be expressed in the basis 
$\left\{ |\uparrow\uparrow\rangle,\;|\uparrow\downarrow\rangle,\;
|\downarrow\uparrow\rangle,\;|\downarrow\downarrow\rangle\right\} $ as 
\begin{equation}
\rho_{ij}=\left(\begin{array}{cccc}
a & 0 & 0 & f\\
0 & b_{1} & z & 0\\
0 & z & b_{2} & 0\\
f & 0 & 0 & d\end{array}\right)
\label{DEN}
\end{equation}
The elements of the reduced density matrix can be expressed in terms 
of the single-site magnetization and two-spin correlation functions. As
shown in Ref. \cite{sarandy}, the eigenvalues of $\rho_{ij}$ are 
\begin{eqnarray}
\lambda_{0} & = & \frac{1}{4}\left\{ \left(1+c_{3}\right)+\sqrt{\left(c_{4}+c_{5}\right)^{2}+\left(c_{1}-c_{2}\right)^{2}}\right\} \nonumber \\
\lambda_{1} & = & \frac{1}{4}\left\{ \left(1+c_{3}\right)-\sqrt{\left(c_{4}+c_{5}\right)^{2}+\left(c_{1}-c_{2}\right)^{2}}\right\} \nonumber \\
\lambda_{2} & = & \frac{1}{4}\left\{ \left(1-c_{3}\right)+\sqrt{\left(c_{4}-c_{5}\right)^{2}+\left(c_{1}+c_{2}\right)^{2}}\right\} \nonumber \\
\lambda_{3} & = & \frac{1}{4}\left\{ \left(1-c_{3}\right)-\sqrt{\left(c_{4}-c_{5}\right)^{2}+\left(c_{1}+c_{2}\right)^{2}}\right\} 
\label{eigen}
\end{eqnarray}
with 
\begin{eqnarray}
c_{1} & = & 2z+2f\nonumber\\
c_{2} & = & 2z-2f\nonumber\\
c_{3} & = & a+d-b_{1}-b_{2}\nonumber\\
c_{4} & = & a-d-b_{1}+b_{2}\nonumber\\
c_{5} & = & a-d+b_{1}-b_{2}
\label{coeff}
\end{eqnarray}
The mutual information (equation \ref{mutual3}) can be written as \cite{sarandy,luo}
\begin{equation}
I\left(\rho_{AB}\right)=S\left(\rho_{A}\right)+S\left(\rho_{B}\right)
+\sum_{\alpha=0}^{3}\lambda_{\alpha}\log_{2}\lambda_{\alpha}
\label{MI}
\end{equation}
where
\begin{eqnarray}
S\left(\rho_{A}\right) & = & -\frac{\left(1+c_{5}\right)}{2}\log_{2}\frac{\left(1+c_{5}\right)}{2}
-\frac{\left(1-c_{5}\right)}{2}\log_{2}\frac{\left(1-c_{5}\right)}{2} \nonumber\\
S\left(\rho_{B}\right) & = & -\frac{\left(1+c_{4}\right)}{2}\log_{2}\frac{\left(1+c_{4}\right)}{2}
-\frac{\left(1-c_{4}\right)}{2}\log_{2}\frac{\left(1-c_{4}\right)}{2}
\label{SI} 
\end{eqnarray}
The reduced density matrix $\rho_{ij}$ (equation \ref{DEN}) has a simpler
form when specific symmetries of the spin Hamiltonian are taken into
account. The element $f=0$ when the $z$-component of the total spin
commutes with the Hamiltonian, i.e., is a conserved quantity. Also,
$a=d$, $b_{1}=b_{2}$ when the magnetization density has expectation
value zero resulting in $c_{4}=c_{5}=0$ and $c_{1}=c_{2}$ (equation \ref{coeff}). 
Under these simplifications, the maximization procedure for
computing the classical correlations $C\left(\rho_{AB}\right)$ (equation \ref{class}) 
can be carried out analytically to yield \cite{sarandy,luo}
\begin{equation}
C\left(\rho_{AB}\right)=\frac{(1-c)}{2}\log_{2}(1-c)+\frac{(1+c)}{2}\log_{2}(1+c)
\label{CI}
\end{equation}
where $c=\max\left(\left|c_{1}\right|,\left|c_{2}\right|,\left|c_{3}\right|\right)$.
The QD, $Q\left(\rho_{AB}\right)$, is given by \cite{sarandy,luo}
\begin{eqnarray}
Q\left(\rho_{AB}\right) & = & I\left(\rho_{AB}\right)-C\left(\rho_{AB}\right) \nonumber\\
& = &  \frac{1}{4}[(1-c_{1}-c_{2}-c_{3})\log_{2}(1-c_{1}-c_{2}-c_{3})
 +(1-c_{1}+c_{2}+c_{3})\log_{2}(1-c_{1}+c_{2}+c_{3}) \nonumber \\
& & +(1+c_{1}-c_{2}+c_{3})\log_{2}(1+c_{1}-c_{2}+c_{3})
 +(1+c_{1}+c_{2}-c_{3})\log_{2}(1+c_{1}+c_{2}-c_{3})] \nonumber \\
& & -\frac{(1-c)}{2}\log_{2}(1-c)-\frac{(1+c)}{2}\log_{2}(1+c)
\label{QI}
\end{eqnarray}

The ground and thermal state entanglement properties of small spin clusters have been computed 
in earlier studies \cite{dowling,bose1,pal}. In this paper, we compute the mutual 
information, classical correlation and 
QD in the ground and thermal states of a symmetric spin trimer and a spin tetramer with 
nearest-neighbour (n.n.) as well as next-nearest-neighbour (n.n.n.) interactions. There are several 
examples of molecular magnets represented by spin trimers and tetramers \cite{haraldsen}. In sections II 
and III, the results for the spin trimer and the tetramer respectively are presented. In section 
IV, we analyze some earlier results \cite{werlang2} on the robustness of QD to sudden death in the context 
of the spin trimer and tetramer. Section V contains a summary of the main results obtained in this
paper and concluding remarks. 

\section{Classical and Quantum Correlations in Spin Trimer}

The symmetric spin trimer consisting of three spins of magnitude $\frac{1}{2}$ is described by the 
Heisenberg exchange interaction Hamiltonian
\begin{eqnarray}
 H_{trimer}=J\sum_{i=1}^{3}S_{i}^{z}S_{i+1}^{z} 
            + \epsilon J \sum_{i=1}^{3}\left(S_{i}^{x}S_{i+1}^{x}+S_{i}^{y}S_{i+1}^{y}\right)
 \label{HTRIMER}
\end{eqnarray}
where $S_{i}^{\alpha}$ $\left(\alpha =x,y,z\right)$ defines the spin operator at the $i$th site 
of the trimer, $\epsilon$ is an anisotropy parameter $(\epsilon \leq1)$ 
and $J$ is the strength of the exchange interaction. The eigenstates of the trimer are 
given by
\begin{eqnarray}
  |1\rangle & = & |\uparrow \uparrow \uparrow \rangle \nonumber \\
 |2\rangle & = & 
 \frac{1}{\sqrt{3}}\left(q|\uparrow\uparrow\downarrow\rangle+q^{2}|\uparrow\downarrow\uparrow\rangle+|\downarrow\uparrow\uparrow\rangle\right) \nonumber \\ 
 |3\rangle & = & 
 \frac{1}{\sqrt{3}}\left(q^{2}|\uparrow\uparrow\downarrow\rangle+q|\uparrow\downarrow\uparrow\rangle+|\downarrow\uparrow\uparrow\rangle\right) \nonumber \\ 
 |4\rangle & = & 
 \frac{1}{\sqrt{3}}\left(|\uparrow\uparrow\downarrow\rangle+|\uparrow\downarrow\uparrow\rangle+|\downarrow\uparrow\uparrow\rangle\right) \nonumber \\ 
 |5\rangle & = & 
 \frac{1}{\sqrt{3}}\left(q|\downarrow\downarrow\uparrow\rangle+q^{2}|\downarrow\uparrow\downarrow\rangle+|\uparrow\downarrow\downarrow\rangle\right) \nonumber \\ 
 |6\rangle & = & 
 \frac{1}{\sqrt{3}}\left(q^{2}|\downarrow\downarrow\uparrow\rangle+q|\downarrow\uparrow\downarrow\rangle+|\uparrow\downarrow\downarrow\rangle\right) \nonumber \\ 
 |7\rangle & = & 
 \frac{1}{\sqrt{3}}\left(|\downarrow\downarrow\uparrow\rangle+|\downarrow\uparrow\downarrow\rangle+|\uparrow\downarrow\downarrow\rangle\right) \nonumber \\ 
 |8\rangle & = & |\downarrow \downarrow \downarrow \rangle
 \label{3STATE}
\end{eqnarray}
where $q=e^{i\frac{2\pi}{3}}$ is the cube root of unity satisfying $q^{3}=1$ and $q+q^{2}+1=0$. The 
corresponding energy eigenvalues are 
\begin{eqnarray}
 E_{1} & = & E_{8}  =  \frac{3J}{4} \nonumber \\
 E_{2} & = & E_{3}=E_{5}  =  E_{6}=-\frac{\left(1+2\epsilon\right)J}{4} \nonumber \\
 E_{4} & = & E_{7}  =  -\frac{\left(1-4\epsilon\right)J}{4}
 \label{3EN}
\end{eqnarray}
We first calculate the ground state ($T=0$) QD in the isotropic case $\epsilon=1$. There are two 
distinct eigenenergies given by 
\begin{eqnarray}
 e_{1} & = & E_{1}=E_{4}=E_{7}=E_{8}=\frac{3J}{4} \nonumber \\
 e_{2} & = & E_{2}=E_{3}=E_{5}=E_{6}=-\frac{3J}{4}
\label{ee}
\end{eqnarray}
In the case of antiferromagnetic (AFM) exchange interaction, $J>0$, the ground state is four-fold
degenerate with the energy $e_{2}$. The ground state density matrix is given by 
\begin{eqnarray}
 \rho_{g}^{AFM}=\frac{1}{4}\left(|2\rangle\langle2|+|3\rangle\langle3|
                                 +|5\rangle\langle5|+|6\rangle\langle6|\right)
\end{eqnarray}
The two-qubit reduced density matrix in the standard basis has the form shown in equation \ref{DEN}
with 
\begin{eqnarray}
a=d=\frac{1}{6},\;b_{1}=b_{2}=\frac{1}{3},\;z=-\frac{1}{6},\;f=0 
\end{eqnarray}
From equation \ref{coeff}, $c=\max \left(\left|c_{1}\right|,\left|c_{2}\right|,\left|c_{3}\right|\right)=\frac{1}{3}$. 
The eigenvalues of the reduced density matrix are (equation \ref{eigen}) with $c_{4}=0,c_{5}=0$):
\begin{eqnarray}
\lambda_{1}=\lambda_{2}=\lambda_{3}=\frac{1}{6},\;\lambda_{4}=\frac{1}{2} 
\end{eqnarray}
From equations (\ref{MI}), (\ref{SI}), (\ref{CI}) and (\ref{QI}), the mutual information $I\left(\rho_{AB}\right)$, the classical correlation 
$C\left(\rho_{AB}\right)$ and the QD, $Q\left(\rho_{AB}\right)$, are:
\begin{eqnarray}
I\left(\rho_{AB}\right)=0.207,\;C\left(\rho_{AB}\right)=0.082,\;Q\left(\rho_{AB}\right)=0.125 
\label{QDAFM}
\end{eqnarray}
We next consider the ferromagnetic (FM) case with $J<0$. The ground state is four-fold degenerate, as in the AFM 
case, with the ground state energy $e_{1}$ (equation \ref{ee}). The ground state density matrix $\rho_{g}^{FM}$ is given 
by 
\begin{eqnarray}
\rho_{g}^{FM} =\frac{1}{4}\left(|1\rangle\langle1|+|4\rangle\langle4|
                                 +|7\rangle\langle7|+|8\rangle\langle8|\right)
\end{eqnarray}
The reduced density matrix has the same form as in the AFM case with 
\begin{eqnarray}
 a=d=\frac{1}{3},\;b_{1}=b_{2}=\frac{1}{6},\;z=\frac{1}{6},\;f=0
\end{eqnarray}
The  value of $c$ is $\frac{1}{3}$ and the eigenvalues of the reduced density matrix are 
\begin{eqnarray}
\lambda_{1}=\lambda_{2}=\lambda_{3}=\frac{1}{3},\;\lambda_{4}=0  
\end{eqnarray}
The values of the mutual information, classical correlation and QD are 
\begin{eqnarray}
I\left(\rho_{AB}\right)=0.415,\;C\left(\rho_{AB}\right)=0.082,\;Q\left(\rho_{AB}\right)=0.333 
\label{QDFM}
\end{eqnarray}
From equations (\ref{QDAFM}) and (\ref{QDFM}), one finds that the magnitude of the classical correlation $C\left(\rho_{AB}\right)$
is the same in the FM and AFM cases whereas the QD has a lower value in the latter case. 

\begin{figure}
 \begin{center}
 \includegraphics[scale=0.6]{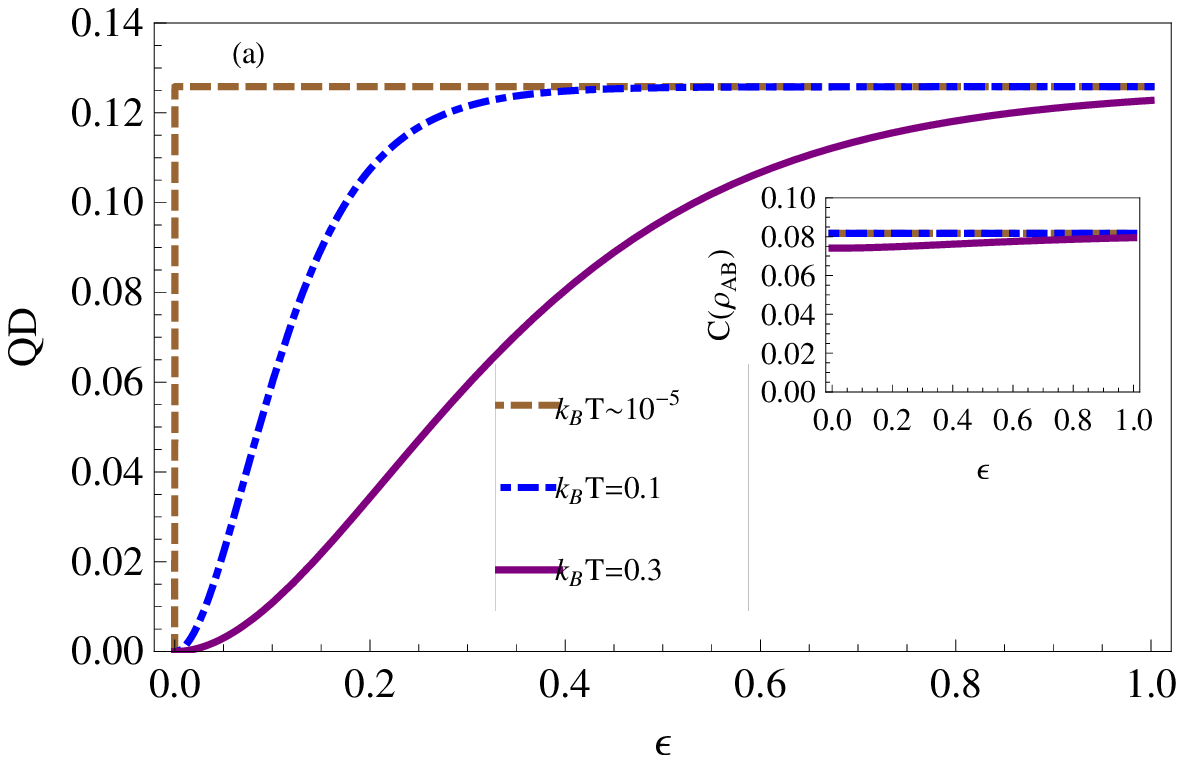}\linebreak
 \includegraphics[scale=0.6]{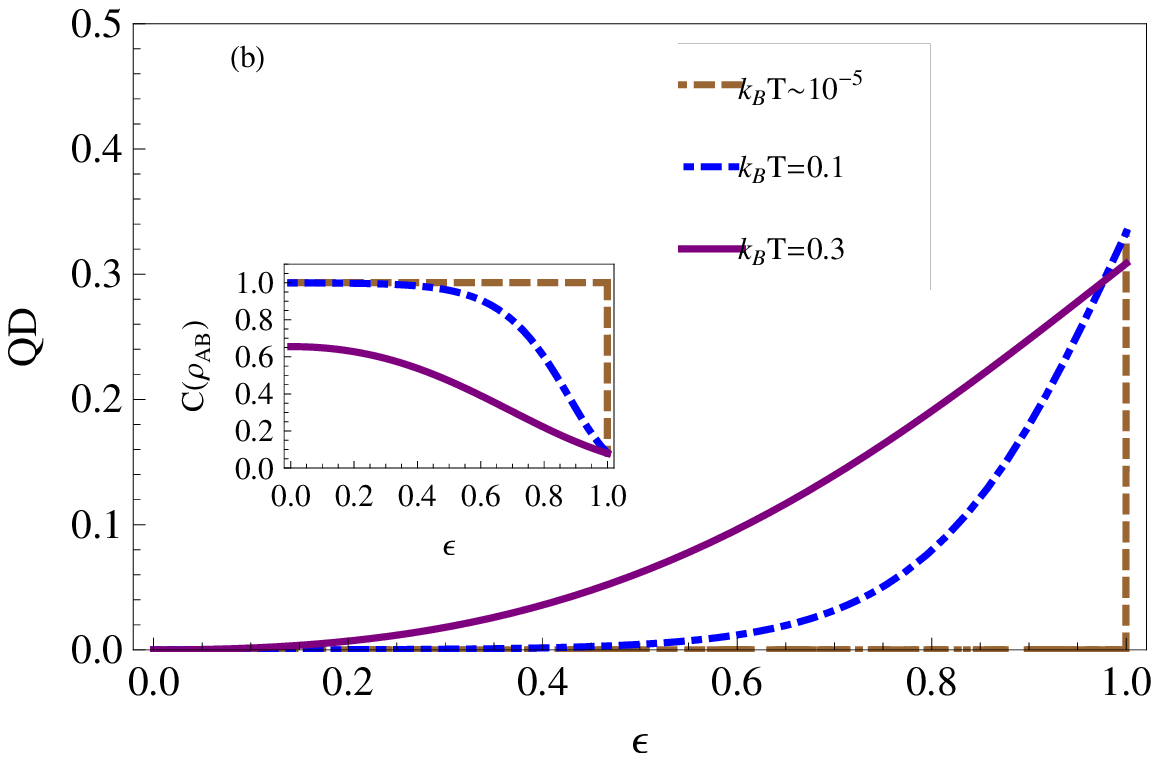}
 \end{center}
 \caption{Variation of quantum discord (QD) and classical correlation $C\left(\rho_{AB}\right)$ (inset) as 
a function of the anisotropy parameter $\epsilon$ for different temperatures in the (a) AFM and (b) FM cases 
with $|J|=1$. }
\label{trimer_fig1}
\end{figure}

We next compute the thermal state QD. From equation \ref{3EN}, the partition function is 
\begin{eqnarray}
 Z=2\left(e^{-3\gamma}+2e^{(1+2\epsilon)\gamma}+e^{(1-4\epsilon)\gamma}\right)
\end{eqnarray}
where $\gamma=\frac{J}{4k_{B}T}$. The thermal state density matrix is 
\begin{eqnarray}
\rho_{T}=\frac{1}{Z}\sum_{i=1}^{8}e^{-\beta E_{i}}|i\rangle\langle i| 
\end{eqnarray}
with $\beta=\frac{1}{k_{B}T}$. The reduced thermal state density matrix has the form shown in equation \ref{DEN} with  
\begin{eqnarray}
a & = & d=\frac{1}{Z}\left(e^{-3\gamma}+\frac{2}{3}e^{(1+2\epsilon)\gamma}
                           +\frac{1}{3}e^{(1-4\epsilon)\gamma}\right) \nonumber \\
b_{1} & = & b_{2}=\frac{2}{3Z}\left(2e^{(1+2\epsilon)\gamma}+e^{(1-4\epsilon)\gamma}\right) \nonumber \\
z & = & \frac{2}{3Z}\left(e^{(1-4\epsilon)\gamma}-e^{(1+2\epsilon)\gamma}\right) \nonumber \\
f & = & 0
\label{DENTRIMER} 
\end{eqnarray}
We calculate $C\left(\rho_{AB}\right)$ and QD as functions of the parameter $\epsilon$ and temperature $T$. 
Figure \ref{trimer_fig1} shows the variation of the QD and $C\left(\rho_{AB}\right)$ (inset) with the anisotropy parameter $\epsilon$. 
In both the AFM and FM cases, the QD increases with $\epsilon$. At a fixed value of $\epsilon$, the QD decreases with 
increasing $T$ in the AFM case though non-monotonic behaviour is observed in the FM case. Furthermore, 
$C\left(\rho_{AB}\right)$ (inset) increases slowly with $\epsilon$ for a fixed value of $T$ in the AFM case whereas 
in the FM case it decreases with $\epsilon$. In both the cases, $C\left(\rho_{AB}\right)$ decreases with $T$ 
for a fixed value of $\epsilon$.

\begin{figure}
 \begin{center}
 \includegraphics[scale=0.6]{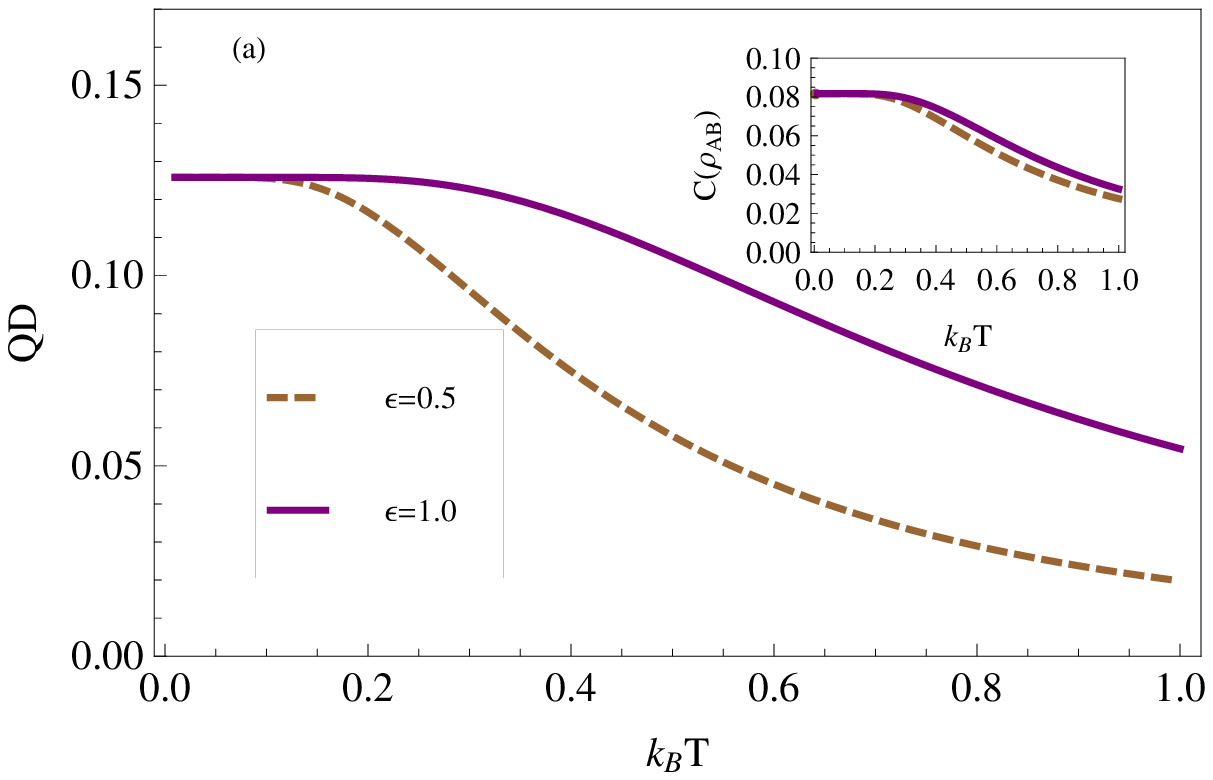}\linebreak
 \includegraphics[scale=0.6]{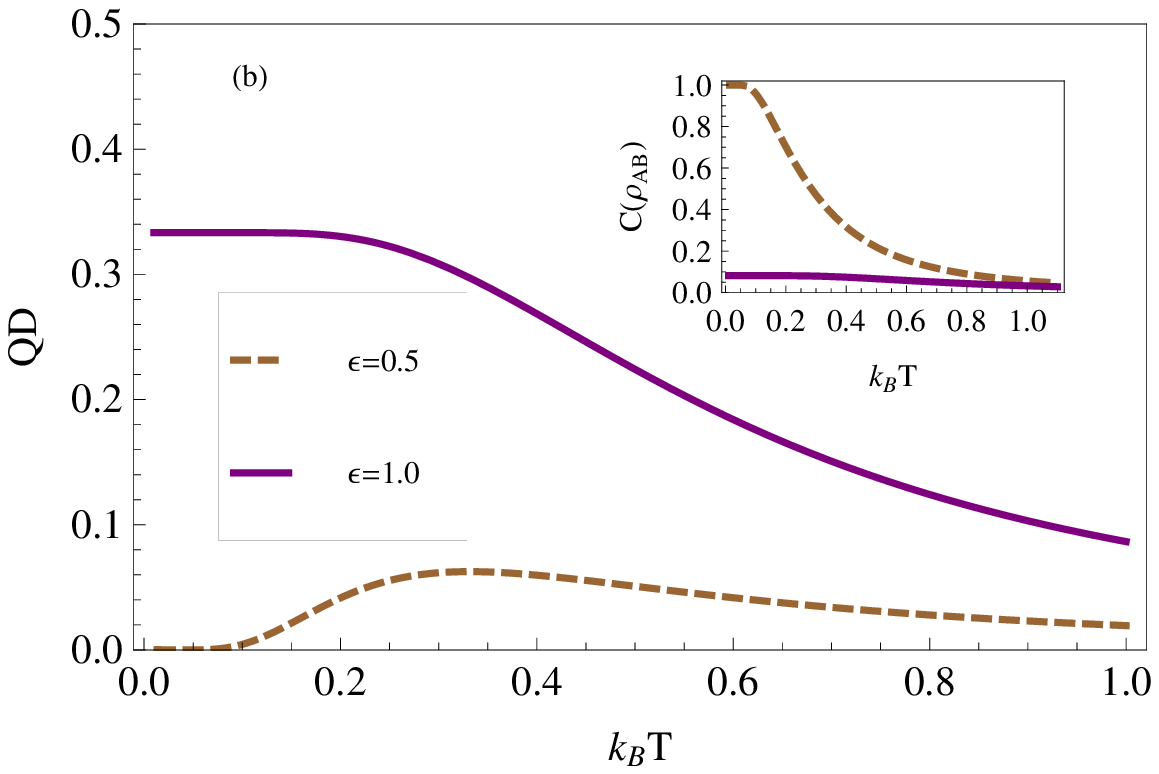}
 \end{center}
 \caption{Variation of quantum discord (QD) and classical correlation $C\left(\rho_{AB}\right)$ (inset) as 
a function of temperature $T$ for different values of the anisotropy parameter $\epsilon$ in the (a) AFM 
and (b) FM cases with $|J|=1$.}
\label{trimer_fig2}
\end{figure}

Figure \ref{trimer_fig2} shows the variation of QD and $C\left(\rho_{AB}\right)$ with temperature for different values of 
$\epsilon$ in both the AFM and FM cases. In the AFM case, the QD decreases with $T$ for both low and high values of 
$\epsilon$. In the FM case, for low values of $\epsilon$, the QD first increases with $T$, reaches a maximum value and 
then decreases. For high values of $\epsilon$, the same features as in the AFM case are observed. 
For the AFM trimer, it is well-known \cite{wang} that there is no pairwise entanglement both at $T=0$ and at 
finite temperatures. For the FM trimer, the same is true when $0<\epsilon\leq 1$. 
The calculations of the present study show that the QD, a different measure of quantum 
correlations, has non-zero  values both in the ground and thermal states. Another remarkable feature is that
though the magnitude of the QD decreases as $T$ increases, it falls to zero only asymptotically, i.e., as 
$T\rightarrow \infty$. This is in contrast to the property of pairwise entanglement in the thermal states of 
spin clusters \cite{bose1,pal,wang}. The concurrence, a measure of the pairwise entanglement, falls to zero 
value above a finite temperature.
A proof that the QD has a non-zero value in  the thermal states of the symmetric trimer 
over an extended temperature region is obtained using the result quoted in Ref. \cite{bylicka}. 
The two-qubit reduced 
density matrix (equation \ref{DEN}) has vanishing discord if the following conditions are satisfied:
\begin{eqnarray}
|f|=|z|,\;a=b_{2},\;d=b_{1}
\label{QDVANISH} 
\end{eqnarray}
The reduced density matrix elements for the spin trimer are given in equation \ref{DENTRIMER} and one can verify that the 
conditions set in equation \ref{QDVANISH} are obeyed only when $T\rightarrow\infty$.         
   
We now briefly consider the case when an external magnetic field term, $h\sum_{i=1}^{3}S_{i}^{z}$, is introduced 
in the spin trimer Hamiltonian (equation \ref{HTRIMER}). The energy eigenvalues are now 
\begin{eqnarray}
 E_{1} & = & \frac{3J}{4}+\frac{3h}{2} \nonumber \\
 E_{2} & = & E_{3}=\frac{h}{2}-\frac{\left(1+2\epsilon\right)J}{4} \nonumber \\
 E_{4} & = & \frac{h}{2}-\frac{\left(1-4\epsilon\right)J}{4} \nonumber \\
 E_{5} & = & E_{6}=-\frac{h}{2}-\frac{\left(1+2\epsilon\right)J}{4} \nonumber \\
 E_{7} & = & -\frac{h}{2}-\frac{\left(1-4\epsilon\right)J}{4} \nonumber \\
 E_{8} & = & \frac{3J}{4}-\frac{3h}{2} 
\end{eqnarray}
One has $a\neq d$ in equation \ref{DEN} so that $c_{4}=c_{5}\neq0$ (equation \ref{coeff}). In this case, QD has to be computed 
numerically. We briefly discuss the results obtained for the case of $\epsilon=1$. In the AFM case, the ground 
state is doubly degenerate for $h<\frac{3J}{2}$ and the ground state density matrix is 
\begin{eqnarray}
\rho_{h}^{g}=\frac{1}{2}\left(|5\rangle\langle5|+|6\rangle\langle6|\right) 
\end{eqnarray}
The elements of the reduced density matrix (equation \ref{DEN}) are
\begin{eqnarray}
a=0,\;d=\frac{1}{3},\;b_{1}=b_{2}=\frac{1}{3},\;z=-\frac{1}{6} 
\end{eqnarray}
The numerically computed value of the QD is 0.125815. For all values of $h>\frac{3J}{2}$, the ground state 
is the fully separable state $|8\rangle$ with QD = 0. At $h=\frac{3J}{2}$, a first order quantum phase transition 
(QPT) takes place and the ground state is three fold degenerate:
\begin{eqnarray}
\rho_{h}^{g,QPT}=\frac{1}{3}\left(|5\rangle\langle5|+|6\rangle\langle6|+|8\rangle\langle8|\right) 
\end{eqnarray}
The reduced density matrix has the elements 
\begin{eqnarray}
a=0,\;d=\frac{5}{9},\;b_{1}=b_{2}=\frac{2}{9},\;z=-\frac{1}{9} 
\end{eqnarray}
\begin{figure}
\begin{center}
\includegraphics[scale=0.6]{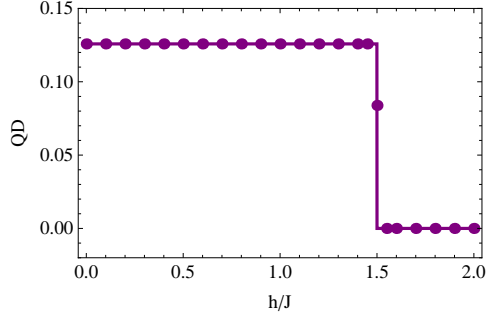}
\end{center}
\caption{Variation of quantum discord (QD) with magnetic field $h$ at $T=0$. The QD remains constant in the range 
$0\leq h\leq \frac{3J}{2}$. At $h=\frac{3J}{2}$, the value of QD is 0.0838764. When $h$ is $>\frac{3J}{2}$, the QD 
is zero as state $|8\rangle$ is the ground state.}
\label{trimer_fig3} 
\end{figure}
\begin{figure}
\begin{center}
\includegraphics[scale=0.6]{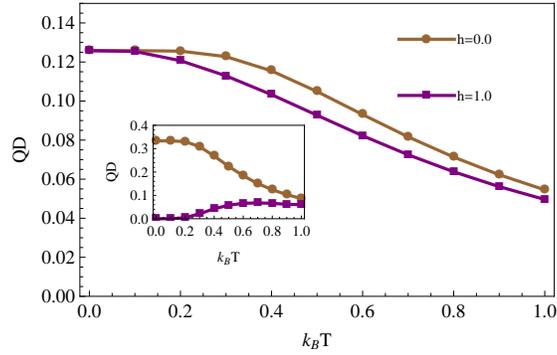}
\end{center}
\caption{Variation of quantum discord (QD) as a function of temperature for different values of $h$ and 
$\epsilon=1.0$ ($|J|=1.0$) in the AFM and FM (inset) cases.}
\label{trimer_fig4} 
\end{figure}
The QD has the value 0.0838764. Figure \ref{trimer_fig3} shows the variation of QD with the magnetic field $h$ at $T=0$.
Figure \ref{trimer_fig4} shows the variation of QD with $h$ in the AFM and the FM cases. For comparison, the plots for $h=0$
are also included. In the AFM case, the effect of $h$ on QD is not prominent. In the FM case, the QD decreases as 
$h$ increases. For $h=1$, the QD is zero at $T=0$, rises with temperature initially and then
assymptotically goes to zero.

\section{Classical and Quantum Correlations in Spin Tetramer}
\begin{figure}
 \begin{center}
 \includegraphics[scale=0.5]{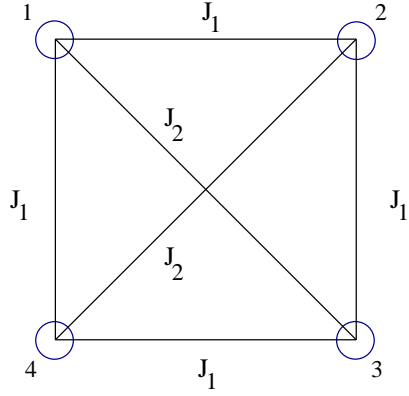}
 \end{center}
 \caption{A tetramer of spins of magnitude $\frac{1}{2}$. $J_{1}$ and $J_{2}$ denote the strengths of the n.n. 
and n.n.n. (diagonal) exchange interactions.}
\label{tetramer}
\end{figure}

\begin{figure}
 \begin{center}
 \includegraphics[scale=0.6]{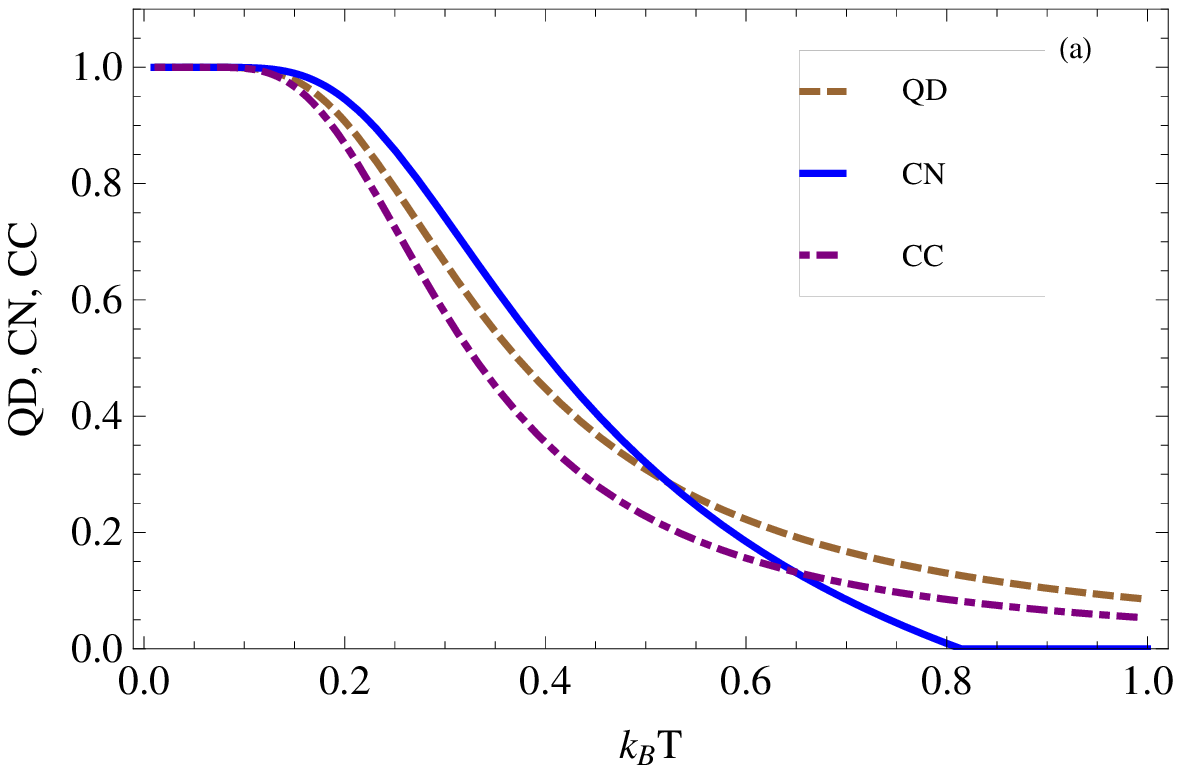}\linebreak
 \includegraphics[scale=0.6]{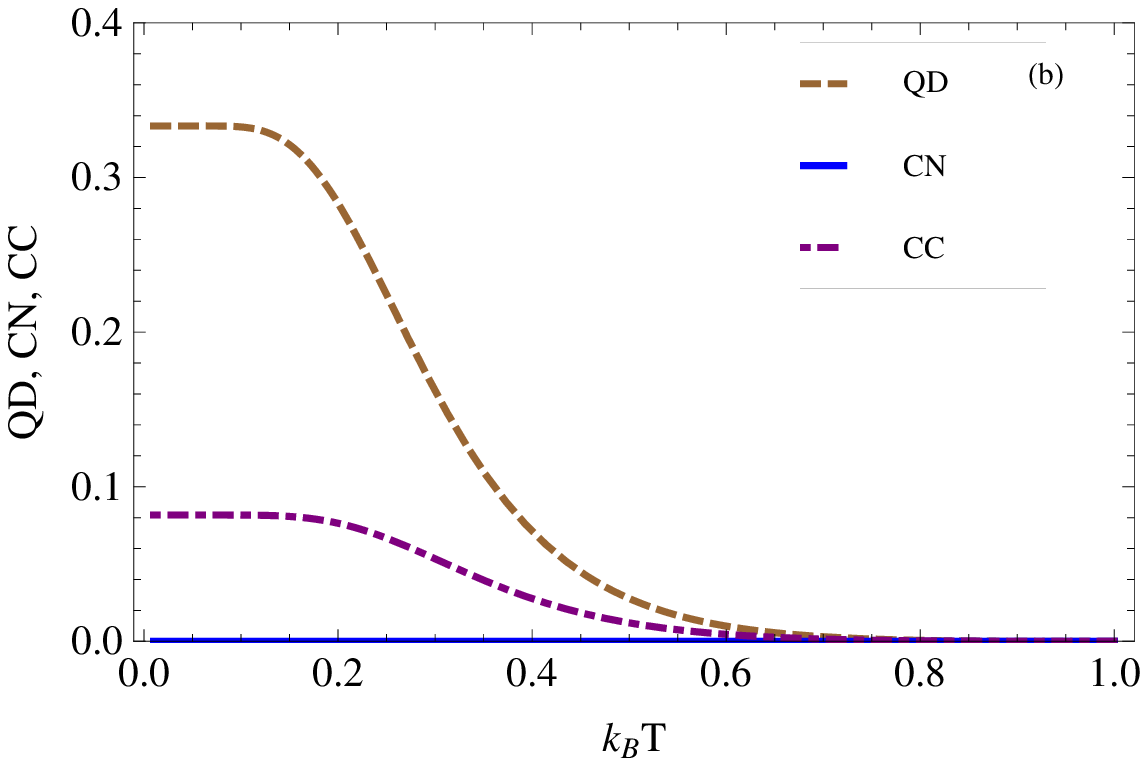}
 \end{center}
 \caption{Variation of quantum discord (QD), concurrence (CN) and classical correlation (CC) as functions 
of temperature for (a) $J_{1}<J_{2}\; (2J_{1}=J_{2}=1.0)$ and (b) $J_{1}>J_{2}\; (J_{1}=2J_{2}=1.0)$.
The results correspond to the n.n.n. reduced density matrix.}
\label{tetra_fig2}
\end{figure}

The spin tetramer cluster is shown in figure \ref{tetramer}. The n.n. spins interact via the isotropic Heisenberg 
exchange interaction Hamiltonian with interaction strength $J_{1}$. The n.n.n. spins interact with exchange 
interaction strength $J_{2}$. We consider the AFM case with $J_{1},J_{2}>0$.
\begin{eqnarray}
H_{tetramer} =  J_{1}\left(\overset{\rightarrow}{S}_{1}.\overset{\rightarrow}{S}_{2}+
                         \overset{\rightarrow}{S}_{2}.\overset{\rightarrow}{S}_{3}+
                         \overset{\rightarrow}{S}_{3}.\overset{\rightarrow}{S}_{4}+
                         \overset{\rightarrow}{S}_{4}.\overset{\rightarrow}{S}_{1}\right)
                         + J_{2}\left(\overset{\rightarrow}{S}_{1}.\overset{\rightarrow}{S}_{3}+
                         \overset{\rightarrow}{S}_{2}.\overset{\rightarrow}{S}_{4}\right)
\label{HTETRAMER}
\end{eqnarray}
The eigenstates of the Hamiltonian are given by 
\begin{eqnarray}
 |1\rangle & = & |\uparrow\uparrow\uparrow\uparrow\rangle \nonumber \\
 |2\rangle & = & \frac{1}{\sqrt{4}}\left(|\uparrow\uparrow\uparrow\downarrow\rangle+i|\uparrow\uparrow\downarrow\uparrow\rangle 
           -|\uparrow\downarrow\uparrow\uparrow\rangle-i|\downarrow\uparrow\uparrow\uparrow\rangle\right) \nonumber \\
 |3\rangle & = & \frac{1}{\sqrt{4}}\left(|\uparrow\uparrow\uparrow\downarrow\rangle-i|\uparrow\uparrow\downarrow\uparrow\rangle 
           -|\uparrow\downarrow\uparrow\uparrow\rangle+i|\downarrow\uparrow\uparrow\uparrow\rangle\right) \nonumber \\
 |4\rangle & = & \frac{1}{\sqrt{4}}\left(|\uparrow\uparrow\uparrow\downarrow\rangle+|\uparrow\uparrow\downarrow\uparrow\rangle 
           +|\uparrow\downarrow\uparrow\uparrow\rangle+|\downarrow\uparrow\uparrow\uparrow\rangle\right) \nonumber \\
 |5\rangle & = & \frac{1}{\sqrt{4}}\left(|\uparrow\uparrow\uparrow\downarrow\rangle-|\uparrow\uparrow\downarrow\uparrow\rangle 
           +|\uparrow\downarrow\uparrow\uparrow\rangle-|\downarrow\uparrow\uparrow\uparrow\rangle\right) \nonumber \\
 |6\rangle & = & \frac{1}{\sqrt{4}}\left(|\uparrow\uparrow\downarrow\downarrow\rangle-|\downarrow\downarrow\uparrow\uparrow\rangle
           +i|\uparrow\downarrow\downarrow\uparrow\rangle-i|\downarrow\uparrow\uparrow\downarrow\rangle\right) \nonumber \\
 |7\rangle & = & \frac{1}{\sqrt{4}}\left(|\uparrow\uparrow\downarrow\downarrow\rangle-|\downarrow\downarrow\uparrow\uparrow\rangle
           -i|\uparrow\downarrow\downarrow\uparrow\rangle+i|\downarrow\uparrow\uparrow\downarrow\rangle\right) \nonumber \\
 |8\rangle & = & \frac{1}{\sqrt{2}}\left(|\uparrow\downarrow\uparrow\downarrow\rangle-|\downarrow\uparrow\downarrow\uparrow\rangle\right) \nonumber \\
 |9\rangle & = & \frac{1}{\sqrt{6}}(|\uparrow\uparrow\downarrow\downarrow\rangle+|\uparrow\downarrow\downarrow\uparrow\rangle
           +|\downarrow\downarrow\uparrow\uparrow\rangle+|\downarrow\uparrow\uparrow\downarrow\rangle 
           +|\uparrow\downarrow\uparrow\downarrow\rangle+|\downarrow\uparrow\downarrow\uparrow\rangle) \nonumber \\
 |10\rangle & = & \frac{1}{\sqrt{4}}\left(|\uparrow\uparrow\downarrow\downarrow\rangle+|\downarrow\downarrow\uparrow\uparrow\rangle
           -|\uparrow\downarrow\downarrow\uparrow\rangle-|\downarrow\uparrow\uparrow\downarrow\rangle\right) \nonumber \\
 |11\rangle & = & \frac{1}{\sqrt{12}}( 2|\uparrow\downarrow\uparrow\downarrow\rangle+2|\downarrow\uparrow\downarrow\uparrow\rangle
           -|\uparrow\uparrow\downarrow\downarrow\rangle-|\uparrow\downarrow\downarrow\uparrow\rangle
           -|\downarrow\downarrow\uparrow\uparrow\rangle-|\downarrow\uparrow\uparrow\downarrow\rangle) \nonumber \\
 |12\rangle & = & \frac{1}{\sqrt{4}}\left(|\downarrow\downarrow\downarrow\uparrow\rangle+i|\downarrow\downarrow\uparrow\downarrow\rangle
           -|\downarrow\uparrow\downarrow\downarrow\rangle-i|\uparrow\downarrow\downarrow\downarrow\rangle\right) \nonumber \\
 |13\rangle & = & \frac{1}{\sqrt{4}}\left(|\downarrow\downarrow\downarrow\uparrow\rangle-i|\downarrow\downarrow\uparrow\downarrow\rangle
           -|\downarrow\uparrow\downarrow\downarrow\rangle+i|\uparrow\downarrow\downarrow\downarrow\rangle\right) \nonumber \\
 |14\rangle & = & \frac{1}{\sqrt{4}}\left(|\downarrow\downarrow\downarrow\uparrow\rangle+|\downarrow\downarrow\uparrow\downarrow\rangle
           +|\downarrow\uparrow\downarrow\downarrow\rangle+|\uparrow\downarrow\downarrow\downarrow\rangle\right) \nonumber \\
 |15\rangle & = & \frac{1}{\sqrt{4}}\left(|\downarrow\downarrow\downarrow\uparrow\rangle-|\downarrow\downarrow\uparrow\downarrow\rangle
           +|\downarrow\uparrow\downarrow\downarrow\rangle-|\uparrow\downarrow\downarrow\downarrow\rangle\right) \nonumber \\
 |16\rangle & = & |\downarrow\downarrow\downarrow\downarrow\rangle 
\label{4STATES}
\end{eqnarray}
The corresponding eigenvalues are
\begin{eqnarray}
 e_{1} & = & E_{1}=E_{4}=E_{9}=E_{14}=E_{16}=J_{1}+\frac{J_{2}}{2} \nonumber \\
 e_{2} & = & E_{2}=E_{3}=E_{6}=E_{7}=E_{12}=E_{13}=-\frac{J_{2}}{2} \nonumber \\
 e_{3} & = & E_{5}=E_{8}=E_{15}=-J_{1}+\frac{J_{2}}{2} \nonumber \\
 e_{4} & = & E_{10}=-\frac{3J_{2}}{2} \nonumber \\
 e_{5} & = & E_{11}=-2J_{1}+\frac{J_{2}}{2} 
\label{4EN}
\end{eqnarray}
We first consider the case $J_{1}>J_{2}$. The ground state is given by $|11\rangle$, which is an example of a 
resonating valance bond (RVB) state \cite{anderson,bose1,christensen}. The RVB state is a linear superposition of two valance bond states
in one of which there is a pair of horizontal valance bonds whereas in the other the valance bonds are vertical. 
A valance bond represents the singlet spin configuration $\frac{1}{\sqrt{2}}
\left(|\uparrow\downarrow\rangle-|\downarrow\uparrow\rangle\right)$. The elements of the n.n. and n.n.n. reduced density matrices 
are
\begin{eqnarray}
 a^{nn} & = & d^{nn}=\frac{1}{12},b_{1}^{nn}=b_{2}^{nn}=\frac{5}{12},z^{nn}=-\frac{1}{3},f^{nn}=0 \nonumber \\
 a^{nnn} & = & d^{nnn}=\frac{1}{3},b_{1}^{nnn}=b_{2}^{nnn}=\frac{1}{6},z^{nnn}=\frac{1}{6},f^{nnn}=0
\end{eqnarray}
The corresponding classical and quantum correlations are 
\begin{eqnarray}
C_{nn}\left(\rho_{AB}\right) & = & 0.350,\;Q_{nn}\left(\rho_{AB}\right)=0.442 \nonumber \\
C_{nnn}\left(\rho_{AB}\right) & = & 0.082,\;Q_{nnn}\left(\rho_{AB}\right)=0.333 
\end{eqnarray}
When $J_{1}$ is $<J_{2}$, the ground state is given by the state $|10\rangle$ which is again a RVB state \cite{bose1}.
The n.n. and the n.n.n. reduced density matrices have the following elements
\begin{eqnarray}
 a^{nn} & = & d^{nn}=\frac{1}{4},b_{1}^{nn}=b_{2}^{nn}=\frac{1}{4},z^{nn}=f^{nn}=0 \nonumber \\
 a^{nnn} & = & d^{nnn}=0,b_{1}^{nnn}=b_{2}^{nnn}=\frac{1}{2},z^{nnn}=-\frac{1}{2},f^{nnn}=0
\end{eqnarray}
The corresponding classical and quantum correlations are 
\begin{eqnarray}
C_{nn}\left(\rho_{AB}\right) & = & Q_{nn}\left(\rho_{AB}\right)=0.0 \nonumber \\
C_{nnn}\left(\rho_{AB}\right) & = & Q_{nnn}\left(\rho_{AB}\right)=1.0  
\end{eqnarray}
At $J_{1}=J_{2}$, a first order QPT takes place with the ground state density matrix given by 
\begin{eqnarray}
 \rho_{g,3}=\frac{1}{2}\left(|10\rangle\langle10|+|11\rangle\langle11|\right)
\end{eqnarray}
The elements of the n.n. and n.n.n. reduced density matrices are 
\begin{eqnarray}
 a=d=\frac{1}{6},\;b_{1}=b_{2}=\frac{1}{3},\;z=-\frac{1}{6}
\end{eqnarray}
The classical and quantum correlations are 
\begin{eqnarray}
C_{nn}\left(\rho_{AB}\right) & = & C_{nnn}\left(\rho_{AB}\right)=0.082 \nonumber \\
Q_{nn}\left(\rho_{AB}\right) & = & Q_{nnn}\left(\rho_{AB}\right)=0.125 
\end{eqnarray}
We next calculate the classical and quantum correlations in the thermal state of the tetramer. 
The partition function of the system is given by 
\begin{eqnarray}
 Z=5e^{-\beta e_{1}}+6e^{-\beta e_{2}}+3e^{-\beta e_{3}}+e^{-\beta e_{4}}+e^{-\beta e_{5}}
\end{eqnarray}
where the energies $e_{i}$'s are as listed in equation \ref{4EN}. The n.n. reduced density matrix 
has the elements
\begin{eqnarray}
a^{nn} & = & d^{nn}=\frac{1}{Z}
\left(\frac{5}{3}e^{-\beta e_{1}}+\frac{3}{2}e^{-\beta e_{2}}+\frac{1}{2}e^{-\beta e_{3}}+
\frac{1}{4}e^{-\beta e_{4}}+\frac{1}{12}e^{-\beta e_{5}}\right) \nonumber \\
b_{1}^{nn} & = & b_{2}^{nn}=\frac{1}{Z}
\left(\frac{5}{6}e^{-\beta e_{1}}+\frac{3}{2}e^{-\beta e_{2}}+e^{-\beta e_{3}}+
\frac{1}{4}e^{-\beta e_{4}}+\frac{5}{12}e^{-\beta e_{5}}\right) \nonumber \\
z^{nn} & = & \frac{1}{Z}
\left(\frac{5}{6}e^{-\beta e_{1}}-\frac{1}{2}e^{-\beta e_{3}}
-\frac{1}{3}e^{-\beta e_{5}}\right) \nonumber \\
f^{nn} & = & 0
\end{eqnarray}
The n.n.n. reduced density matrix has the elements 
\begin{eqnarray}
a^{nnn} & = & d^{nnn}=\frac{1}{Z}
\left(\frac{5}{3}e^{-\beta e_{1}}+e^{-\beta e_{2}}+e^{-\beta e_{3}}
+\frac{1}{3}e^{-\beta e_{5}}\right) \nonumber \\
b_{1}^{nnn} & = & b_{2}^{nnn}=\frac{1}{Z}
\left(\frac{5}{6}e^{-\beta e_{1}}+2e^{-\beta e_{2}}+\frac{1}{2}e^{-\beta e_{3}}
+\frac{1}{2}e^{-\beta e_{4}}+\frac{1}{6}e^{-\beta e_{5}}\right) \nonumber \\
z^{nn} & = & \frac{1}{Z}
\left(\frac{5}{6}e^{-\beta e_{1}}-e^{-\beta e_{2}}+\frac{1}{2}e^{-\beta e_{3}}
-\frac{1}{2}e^{-\beta e_{4}}+\frac{1}{6}e^{-\beta e_{5}}\right) \nonumber \\
f^{nn} & = & 0 
\end{eqnarray}
\begin{figure}
 \begin{center}
 \includegraphics[scale=0.6]{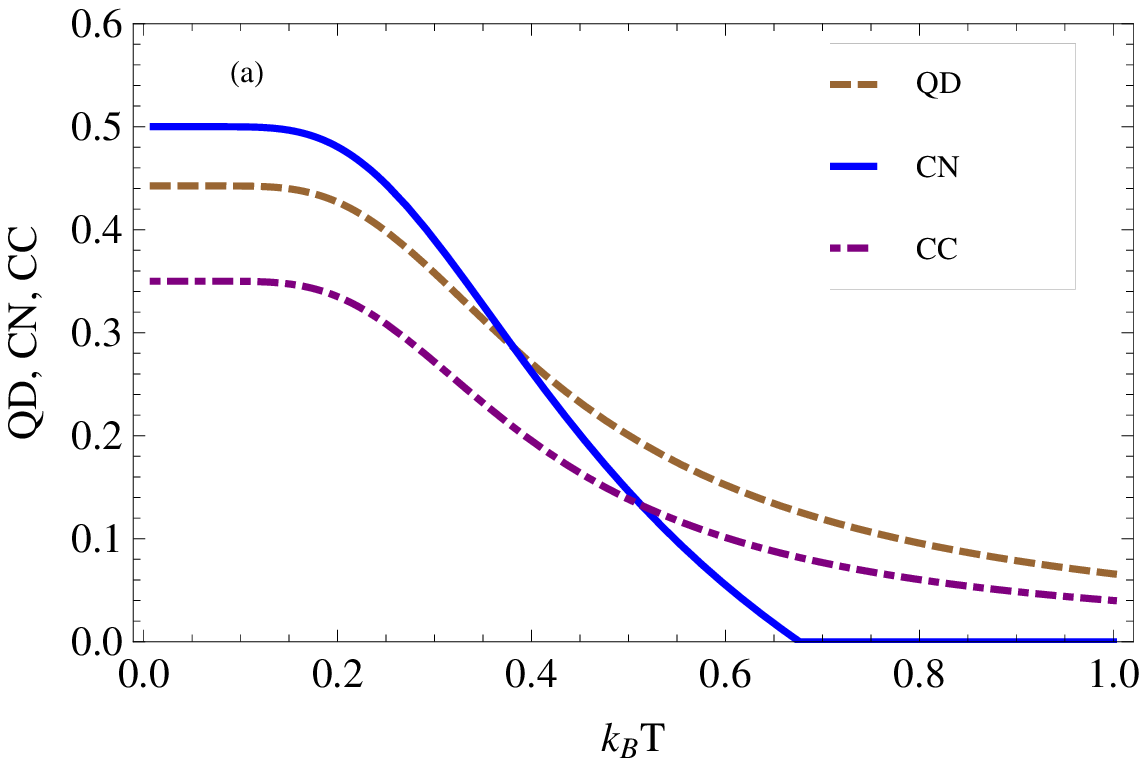}\linebreak
 \includegraphics[scale=0.6]{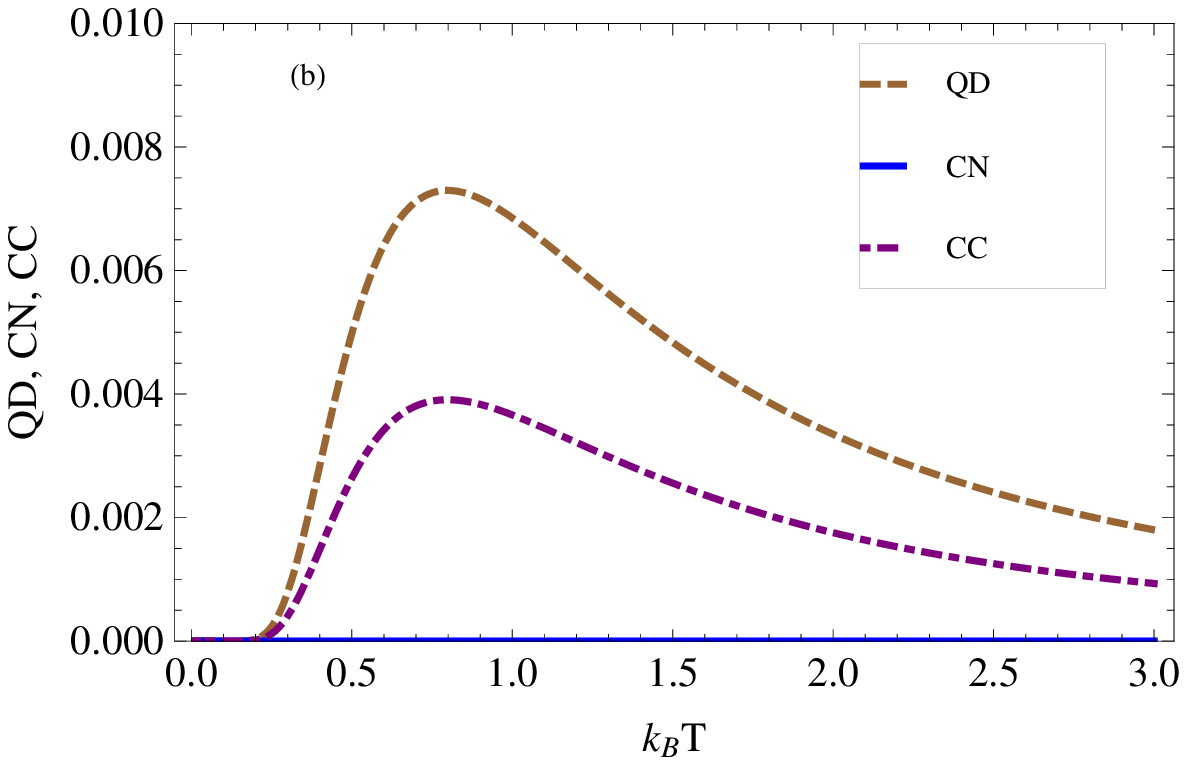}
 \end{center}
 \caption{Variation of quantum discord (QD), concurrence (CN) and classical correlation (CC) as functions 
of temperature for (a) $J_{1}>J_{2}\; (J_{1}=2J_{2}=1.0)$ and (b) $J_{1}<J_{2}\; (2J_{1}=J_{2}=1.0)$. 
The results correspond to the n.n. reduced density matrix.}
\label{tetra_fig1}
\end{figure}
Figure \ref{tetra_fig1} shows the variation of n.n. concurrence (CN), QD and classical correlation 
(CC) with temperature for the AFM case. When $J_{1}>J_{2}$, both the CN and QD decrease with 
temperature but QD has non-zero values at temperatures much higher than the value at which CN 
becomes zero. When $J_{1}<J_{2}$, CN has zero value at all temperatures. The QD is zero at $T=0$,
then it increases with temperature to reach a maximum value after which it decreases with 
temperature. Figure \ref{tetra_fig2} shows the variation of the n.n.n. CN, QD and CC with temperature 
in the AFM case. The CN has non-zero values only when $J_{1}<J_{2}$. The magnitude of CN 
decreases with temperature and falls to zero value at a specific temperature. As in the n.n. 
case, the QD has non-zero values at much higher temperatures.    

\section{Quantum Correlations under Decoherence}
The interaction of a quantum system with its environment results in decoherence, i.e., a 
destruction of the quantum properties including correlations of the system. The dynamics 
of the two-qubit QD under different types of environment have recently been investigated
\cite{werlang2,maziero2,maziero3,mazzola}. In the case of local environments, each qubit interacts with its individual environment.
The channel representing the interaction between a qubit and its environment can be of various 
types: amplitude damping, dephasing, bit flip, phase flip etc. \cite{nielsen}. In this section, we study 
the dynamics of the two-qubit entanglement and QD under the influence of a dephasing channel.
The initial (time $t=0$) two-qubit state is the Werner state
\begin{eqnarray}
 \rho(0)=(1-\alpha)\frac{I}{4}+\alpha|\psi^{-}\rangle\langle\psi^{-}|
\label{wernerstate}
\end{eqnarray}
with $\alpha\in[0,1]$, $I$ the identity matrix and $|\psi^{-}\rangle=\frac{1}{\sqrt{2}}
\left(|\uparrow\downarrow\rangle-|\downarrow\uparrow\rangle\right)$. We note that the two-qubit 
reduced density matrices of the ground states of the AFM spin trimer and tetramer represent 
the Werner states with $\alpha=\frac{1}{3}$ (trimer) and $\alpha=\frac{2}{3}$ (tetramer, 
$J_{1}>J_{2}$) respectively. The dynamics of a two-qubit state under the effect of the dephasing channel 
and with the Werner state as the initial state have been studied in Ref.\cite{werlang2}. We utilize 
the results of this study (some minor errors have been corrected) to examine the evolution of 
the QD in the two-qubit states with $\alpha=\frac{1}{3}$ and $\frac{2}{3}$.

In the Kraus operator representation, an initial state, $\rho(0)$, of the qubits evolves as \cite{werlang2} 
\begin{eqnarray}
\rho(t)=\sum_{\mu ,\nu}E_{\mu , \nu}\rho(0)E_{\mu , \nu}^{\dagger} 
\end{eqnarray}
where the Kraus operators $E_{\mu , \nu}=E_{\mu}\otimes E_{\nu}$ satisfy the completeness relation 
$\sum_{\mu ,\nu}E_{\mu , \nu}E_{\mu , \nu}^{\dagger}=I$ for all $t$. In the case of the dephasing channel, 
the Kraus operators have the matrix form 
\begin{eqnarray}
E_{0}=\left(\begin{array}{cc}
1 & 0 \\
0 & \sqrt{1-\gamma}\end{array}\right),\;
E_{1}=\left(\begin{array}{cc}
0 & 0 \\
0 & \sqrt{\gamma}\end{array}\right)
\label{KRAUS} 
\end{eqnarray}
where $\gamma=1-e^{-\Gamma t}$ with $\Gamma$ denoting the decay rate. The elements of the density 
matrix of the two-qubit system, with $\rho(0)$ given by equation \ref{wernerstate} evolve to 
\begin{eqnarray}
\rho_{ii}(t)=\rho_{ii}(0),\;i=1,...,4 
\end{eqnarray}
\begin{eqnarray}
\rho_{23}(t)=\rho_{23}(0)(1-\gamma)=\rho_{32}(t) 
\end{eqnarray}
The concurrence for the two-qubit evolved state is 
\begin{eqnarray}
CN\left(\rho_{AB}\right)=\alpha\left(\frac{3}{2}-\gamma\right)-\frac{1}{2}
\label{CONCT} 
\end{eqnarray}
The QD for the state is given by 
\begin{eqnarray}
Q\left(\rho_{AB}\right)=\frac{1}{4}\left\{F(a+b)+F(a-b)\right\}-\frac{F(a)}{2}
\label{QDT} 
\end{eqnarray}
\begin{figure}
\begin{center} 
\includegraphics[scale=0.6]{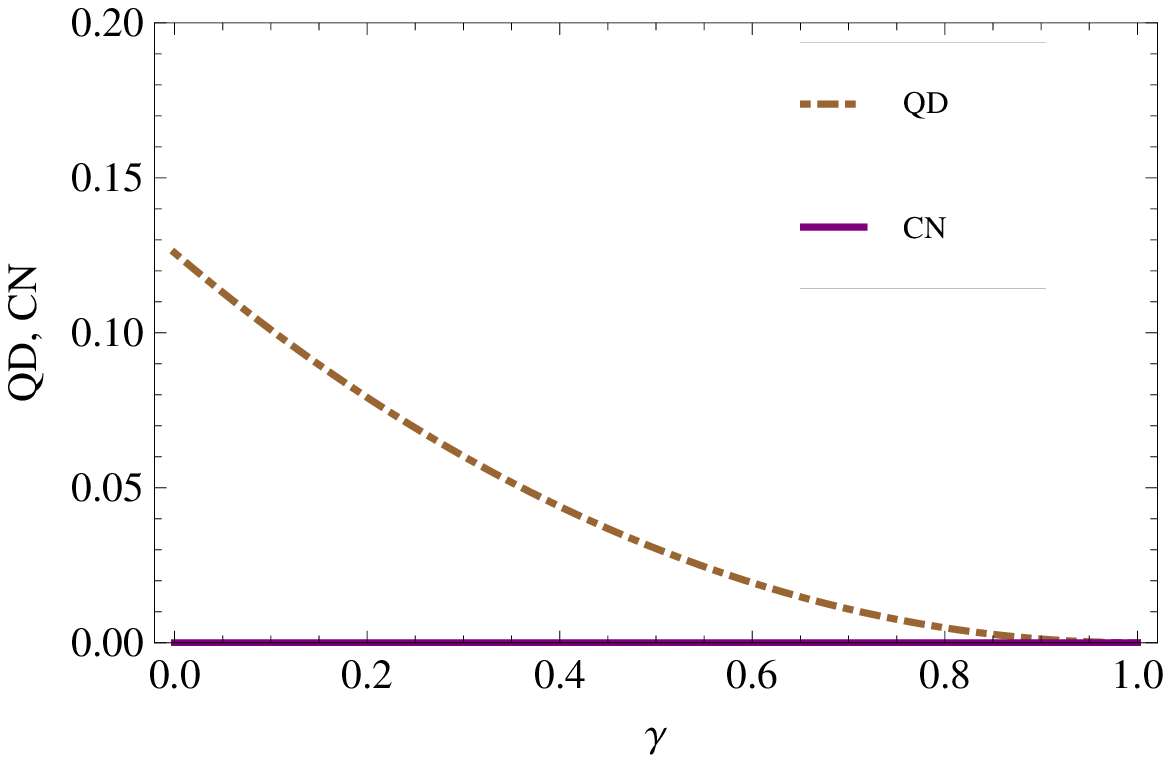}
\end{center}
\caption{Variation of quantum discord (QD) and concurrence (CN) with $\gamma$ for $\alpha=\frac{1}{3}$ 
(equations (\ref{CONCT}) and (\ref{QDT})). The two-qubit reduced density matrix is obtained from the AFM ground state 
of the trimer.} 
\label{decoh1}
\end{figure}
\begin{figure}
\begin{center}
\includegraphics[scale=0.6]{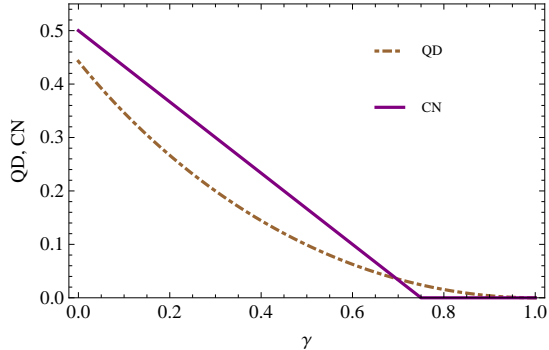}
\end{center}
\caption{Variation of quantum discord (QD) and concurrence (CN) with $\gamma$ for $\alpha=\frac{2}{3}$ 
(equations (\ref{CONCT}) and (\ref{QDT})). The two-qubit reduced density matrix is obtained from the AFM ground state 
of the tetramer.}
\label{decoh2} 
\end{figure}
with $F(x)=x\log_{2}x$, $a=(1+\alpha)$ and $b=2\alpha(1-\gamma)$. Figure \ref{decoh1} (\ref{decoh2})
shows the variation of $CN\left(\rho_{AB}\right)$ and $Q\left(\rho_{AB}\right)$ as a function of 
$\gamma$ for $\alpha=\frac{1}{3}\left(\frac{2}{3}\right)$. In the case of the tetramer, the 
concurrence becomes
zero in a finite time whereas the QD vanishes only in the asymptotic limit $t\rightarrow \infty$.
Thus, the QD, unlike quantum entanglement, exhibits robustness to sudden death.

\section{Discussion}
In this paper, we have calculated the QD of the two-qubit ground and thermal states of a symmetric spin 
trimer and a tetramer with both n.n. and n.n.n. exchange interactions. In both the cases, the QD can be 
evaluated analytically because of the simple structure of the two-qubit reduced density matrix. 
A well-known result pertaining to the spin trimer is that there is no pairwise entanglement 
at both $T=0$ and at finite temperatures for $0<\epsilon\leq 1$ \cite{wang}. We have now shown that the QD has a non-zero value 
for both $T=0$ and $T\neq 0$. An interesting observation is that the QD has a larger value, 
$Q\left(\rho_{AB}\right)=0.333$, in the FM case than that, $Q\left(\rho_{AB}\right)=0.125$, in 
the AFM case. The classical correlation, $C\left(\rho_{AB}\right)$, has the same value 0.082 in 
both the cases with magnitude lower than that of the QD. On the inclusion of an external magnetic field, the QD jumps in magnitude 
at the first-order QPT point, $h_{c}=\frac{3J}{2}$ (figure \ref{trimer_fig3}). This feature is similar to the entanglement jumps 
seen at first-order QPT points \cite{bose2,wu}. Dillenschneider \cite{dillenschneider} has investigated quantum
phase transitions in the one-dimensional spin-$\frac{1}{2}$ transverse Ising and AFM XXZ models using the 
QD as a measure. For both the spin models, the QD displays behaviour similar to that of 
entanglement quantified by the concurrence in the vicinity of the QPT point. In the case of the 
AFM spin tetramer, a first order QPT takes place at $J_{1}=J_{2}$ separating the RVB ground states
$|11\rangle,\;\left(J_{1}>J_{2}\right)$ and  $|10\rangle,\;\left(J_{1}<J_{2}\right)$. Again 
the QD (as well as the classical correlation) exhibits a discontinuity at the transition point. 
This is true in both the cases of n.n. and n.n.n. qubits. A first order QPT involves a 
discontinuity in the first derivative of the ground state energy with respect to a coupling parameter
($h$ in figure \ref{trimer_fig3}). This implies a discontinuity in one or more of the elements of the 
reduced density matrix at the transition point \cite{wu,dillenschneider}. Since the QD is dependent on the elements of the 
reduced density matrix, a first order QPT gives rise to a discontinuity in QD. 
In both the cases of the trimer and the tetramer, 
the asymptotic decay of the QD with temperature indicates that thermal fluctuations cannot kill 
the quantum correlations though the QD is reduced in magnitude at higher temperatures.
Some recent studies on the thermal QD in spin models \cite{werlang1,chen} arrive at a similar 
conclusion.

The successful implementation of quantum computation and communication protocols depends on the 
robustness of entanglement in quantum states. The inevitable interaction between a system and 
its environment results in decoherence and degradation of the  entanglement. The entanglement 
dynamics due to decoherence may bring about the complete disappearance of entanglement at a finite 
time, termed the \textquotedblleft entanglement sudden death\textquotedblright \cite{almeida}. 
Some recent studies \cite{werlang2,mazzola,ferraro} have shown that the QD, in the presence of a Markovian environment 
(memoryless dynamics), decays in time but vanishes only asymptotically. In fact, Ref. \cite{mazzola}
discusses an interesting example of the QD remaining constant up to a time $t=\tilde{t}$ with the 
decay setting in only when $t$ is $>\tilde{t}$. Our studies of the AFM trimer and the tetramer
show that the two-qubit reduced density matrices at $T=0$ have the form of Werner states. As shown 
by Werlang et al. \cite{werlang1}, the QD vanishes asymptotically with time when an initial Werner state is
subjected to a dephasing channel. We have demonstrated this for the trimer and the 
tetramer in Figures (\ref{decoh1}) and (\ref{decoh2}) with the $\gamma\rightarrow 1$ 
$\left(\gamma=1-e^{-\Gamma t}\right)$ limit corresponding to $t\rightarrow \infty$.  The same figures 
show that the pairwise entanglement, as measured by concurrence, is either zero at all times 
(figure \ref{decoh1}) or undergoes a \textquotedblleft sudden death\textquotedblright  at a 
finite time (figure \ref{decoh2}). Some recent studies \cite{datta,lanyon} have shown that the use of states for which 
entanglement is zero (mixed separable states) but QD is non-zero, can improve the efficiency 
of certain computational tasks in comparison with classical computing. The spin trimer ground 
state provides an example of a state with zero entanglement and non-zero QD. Ref. \cite{haraldsen}
provides a number of examples of molecular magnets described by spin trimers and tetramers. 
Molecular spin clusters are ideal candidate systems to test quantum information theoretic concepts. 
Recent advances in supramolecular chemistry provide tools to engineer synthetic spin clusters like a 
molecular cluster of three qubits \cite{timco}. 
The role of decoherence causing crossover from the quantum to the classical domain may be 
ideally studied in mesoscopic systems like molecular spin clusters \cite{troiani}. Since quantum correlations, 
in terms of the QD, persist up to very high temperatures and the decoherence time under specific 
conditions is quite long, such correlations could provide the basis for the implementation 
of quantum information tasks. Molecular magnets, described by small spin clusters, are expected 
to play an important role in such applications. 

In this paper, we have not investigated the dynamics of the QD and the entanglement of the 
reduced two qubit systems subjected to a non-Markovian environment. A number of recent 
studies \cite{wang2,altintas,fanchini,mazzola2} have identified some interesting features of the 
non-Markovian dynamics of the QD which are absent in the Markovian case. The two qubits interact 
with either independent or common non-Markovian environments (reservoirs). In the studies carried 
out so far, the two-qubits are not coupled to each other. For independent reservoirs, the QD is 
found to vanish only at discrete time points \cite{wang2,altintas,fanchini} whereas the 
entanglement disappears in a finite time interval. In the case of a common reservoir, the 
entanglement dynamics exhibit damped oscillations whereas the QD is characterised by isolated 
kinks at which there is a jump in its derivatives \cite{fanchini}. Due to the memory effect 
of the environment, some of the quantum correlations lost during the dissipative dynamics 
can be restored to the qubits giving rise to the \textquotedblleft sudden birth of 
entanglement\textquotedblright and \textquotedblleft revival\textquotedblright of the 
QD \cite{wang2,altintas,fanchini,mazzola2}. As in the Markovian case \cite{mazzola}, for a 
specific class of initial states, the qubit system exhibits a sudden transition between clssical 
and quantum decoherence at time $t=\tilde{t}$ \cite{mazzola2}. For $t < \tilde{t}$, the amount 
of classical correlations decays whereas the QD remains frozen. The reverse situation holds true 
for $t>\tilde{t}$. In the non-Markovian case, multiple such transitions can occur due to the memory 
effect of the environment. For a system of two coupled qubits in a non-Markovian environment, 
there is no study  as yet which investigates the dynamics of the entanglement and the QD in the 
same framework. The spin clusters considered in the present paper correspond to interacting qubit 
systems. Studies on the time evolution of quantum correlation in such systems interacting with 
a non-Markovian environment are essential for a fuller understanding of the problem of decoherence in 
molecular magnets.

\end{document}